\documentclass[aps,prb,twocolumn,preprintnumbers,
amsmath,amssymb,groupedaddress,
superscriptaddress,longbibliography]{revtex4-2}

\usepackage{graphicx}
\usepackage{dcolumn}
\usepackage{bm}%
\usepackage[colorlinks=true,citecolor=blue]{hyperref}
\hypersetup{colorlinks=true,citecolor=blue,linkcolor=red,urlcolor=blue}

\usepackage[T1]{fontenc}
\usepackage{float}

\usepackage{color}
\definecolor{Green}{RGB}{0,204,102}
\definecolor{Purple}{RGB}{102,0,255}
\definecolor{Blue}{RGB}{51,153,255}
\definecolor{Red}{RGB}{151,010,010}

\begin{document}

\sloppy

\title{Dynamical renormalization of electron-phonon coupling in conventional superconductors}

\newcommand*{\DIPC}[0]{{
Donostia International Physics Center (DIPC), 20018 Donostia-San Sebasti\'an, Spain}}

\newcommand*{\IFS}[0]{{Centre for Advanced Laser Techniques, Institute of Physics, 10000 Zagreb, Croatia}}

\author{Nina Girotto}
\affiliation{\IFS}

\author{Dino Novko}
\email{dino.novko@gmail.com}
\affiliation{\IFS}


\begin{abstract}
The adiabatic Born-Oppenheimer approximation is considered to be a robust approach that very rarely breaks down. Consequently, it is predominantly utilized to address various electron-phonon properties in condensed matter physics. By combining many-body perturbation and density functional theories we demonstrate the importance of dynamical (nonadiabatic) effects in estimating superconducting properties in various bulk and two-dimensional materials.
Apart from the expected long-wavelength nonadiabatic effects, we found sizable nonadiabatic Kohn anomalies away from the Brillouin zone center for materials with strong intervalley electron-phonon scatterings.
Compared to the adiabatic result, these dynamical phonon anomalies can significantly modify electron-phonon coupling strength $\lambda$ and superconducting transition temperature $T_c$. Further, the dynamically-induced modifications of $\lambda$ have a strong impact on transport properties, where probably the most interesting is the rescaling of the low-temperature and low-frequency regime of the scattering time $1/\tau$ from about $T^3$ to about $T^2$, resembling the Fermi liquid result for electron-electron scattering.
Our goal is to point out the potential implications of these nonadiabatic effects and reestablish their pivotal role in computational estimations of electron-phonon properties.
\end{abstract}

\maketitle

\section{Introduction}

Electron-phonon coupling (EPC) is crucial for understanding a vast number of phenomena in condensed matter physics, including resistivity, optical absorption, band gap renormalizations, structural phase transitions, charge density waves (CDW), and superconductivity\,\cite{giustino17}. Most of the theoretical considerations of these properties rely on the adiabatic Born-Oppenheimer approximation, where electron and lattice degrees of freedom are treated separately and the dynamical effects of EPC are absent\,\cite{baroni01}. However, there are certain conditions under which the adiabatic approximation fails. Specifically, provided that the electronic dampings are negligible, optical phonons around the center of the Brillouin zone (i.e., $\mathbf{q\cdot v_F}<\omega$ and $1/\tau\ll \omega$) are expected to be affected by the nonadiabatic (NA) corrections \,\cite{migdal58,engelsberg63,maksimov96}. The latter was confirmed by various theoretical and experimental (e.g., Raman and inelastic x-ray scattering) works, where profound NA renormalizations of optical phonons were found, for instance, in doped bulk semiconductors\,\cite{cerdeira72}, MgB$_2$\,\cite{cappelluti06,ponosov17,novko2018a,novko20b,cappelluti22}, transition metals\,\cite{ponosov98,ponosov16}, graphene-based materials\,\cite{lazzeri06,pisana07,piscanes07,saitta08}, hole-doped diamond\,\cite{caruso17}, and doped transition metal dichalcogenides\,\cite{sohier19,novko2020a,eiguren20}.

Besides these studies, where the focus is on long-wavelength optical phonons and several others where adiabatic breakdown of electron band structure renormalizations in infrared-active materials is analyzed\,\cite{ponce15,allen17,miglio20}, there are very few quantitative studies going beyond and exploring self-consistently dynamical corrections of electron-phonon properties in realistic materials, like the coupling strengths, superconductivity, and resistivity\,\cite{meng22a,meng22b}. Namely, it was shown that the dynamical screening of phonons, provided by the combination of molecular dynamics and real-time time-dependent-density-functional theory, can lead to significant modifications of EPC strength $\lambda$ and superconductivity transition temperature $T_c$\,\cite{meng22a}, as well as, potential energy surfaces and anharmonicity\,\cite{meng22b}. In addition, important insights were provided by the model calculations (e.g., Holstein Hamiltonian), where the influence of the dynamical phonon renormalization\,\cite{allen74,marsiglio90,nosarzewski21,setty20,setty22} and vertex corrections\,\cite{grimaldi95,cappelluti00,gorkov16} on superconductivity was demonstrated. For example, it was shown that replacing the bare (adiabatic) phonon propagator with the renormalized one (i.e., including phonon self-energy corrections) in Migdal-Eliashberg equations improves considerably its accuracy\,\cite{marsiglio90,nosarzewski21}. Also, accounting for phonon self-energy corrections, damping in particular, was found to be instrumental for describing the superconducting dome structure in ferroelectric materials\,\cite{setty22}.

Notwithstanding, the majority of the theoretical first-principles studies of superconductivity are done in the framework of the adiabatic approximation\,\cite{giustino17}, and the role of NA corrections is still not fully established.

Here we provide a detailed \emph{ab initio} study on the NA phonon renormalizations and the corresponding impact on $\lambda$, $T_c$, as well as electron scattering rates $1/\tau$ relevant in transport and optical absorption. We use density functional perturbation theory (DFPT)\,\cite{baroni01} and NA phonon self-energies\,\cite{giustino17,epw} in order to simulate NA corrections to phonon frequencies and linewidths, which in turn are utilized to renormalize electron-phonon properties. We investigate the NA effects in several relevant bulk and two-dimensional (2D) systems, where conventional, phonon-mediated superconductivity was confirmed by both experiments and theory or just in theory, i.e., MgB$_2$\,\cite{nagamatsu01,bohnen01,kortus01,liu01,yildrim01,choi02,eiguren08,calandra10,margine13}, hole-doped diamond (C)\,\cite{ekimov04,boeri04,xiang04,giustino07,giustino07b}, and doped monolayers: graphene (1L Gr)\,\cite{margine14,ludbrook15,ichinokura16}, graphane (1L Gr-H)\,\cite{savini10}, molybdenum disulfide (1L MoS$_2$)\,\cite{ye12,ge13,rosner14,costanzo16,fu17,piatti18,goiricelaya19}, arsenene (1L As)\,\cite{kong18}, indium selenide (1L InSe)\,\cite{lugovskoi19,alidoosti21}, and wolfram ditelluride (1L WTe$_2$)\,\cite{ebrahim18,yang20}. Contrary to the common belief, our results show that the NA frequency renormalization can be significant also away from the long-wavelength $\mathbf{q}\approx 0$ region, especially in the multiband and multivalley systems such as MgB$_2$, MoS$_2$, As, and WTe$_2$, where strong interband or intervalley electron-phonon scatterings are possible, forming Kohn anomalies. Further, compared to the adiabatic calculations, we obtained considerable modifications of $\lambda$ and $T_c$ once the NA phonon frequency renormalizations and phonon linewidths due to EPC are taken into account. Specifically, the relative changes of $\lambda$, when going from adiabatic to NA results, can range from small (e.g., $3\%$ for MgB$_2$) to considerable (e.g., $42\%$ for 1L InSe). The obtained modifications for $T_c$ are even more dramatic, varying from $10\%$ in the case of MgB$_2$ to about $80\%$ for 1L MoS$_2$ and 1L InSe. Interestingly, in most cases, the NA effects lead to the reduction of $\lambda$, while for MgB$_2$ and C the inclusion of the dynamical phonon linewidths into the calculations of EPC results its increase. Finally, we show how the dynamical electron-phonon effects can significantly influence, both qualitatively and quantitatively, the functional dependence of the electron-hole pair scattering rate $1/\tau_{\rm op}(\omega,T)$ and mass enhancement factor $\lambda_{\rm op}(\omega,T)$\,\cite{shulga91,marsiglio08} on frequency $\omega$ and temperature $T$. For instance, the inclusion of the NA effects modifies the well-known behaviour of $1/\tau_{\rm op}\propto T^3$ to $1/\tau_{\rm op}\propto T^2$\,\cite{fenton69,allen74}, resembling the Fermi liquid result for electron-electron scatterings. The latter might provide further insights into enigmatic $T^2$ resistivity observed in a number of complex materials\,\cite{marel11,lin15,barisic13,stemmer18,wang20,mirjolet21,kamran22}.

\begin{figure}[!t]
\includegraphics[width=0.45\textwidth]{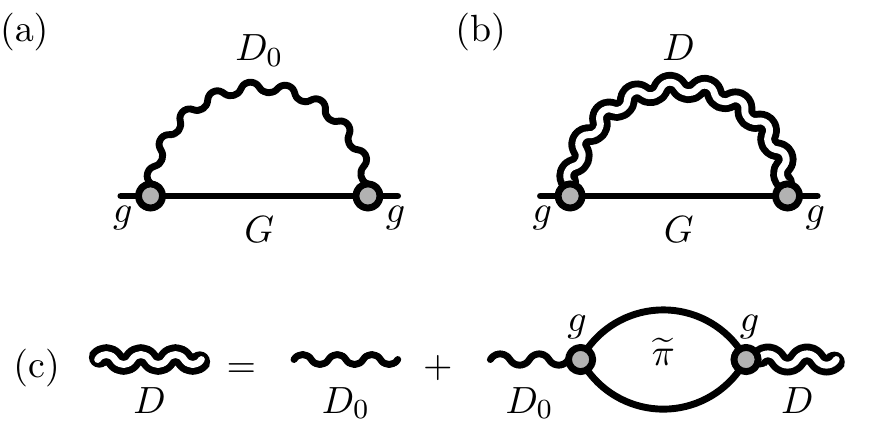}
\caption{ (a) Diagrammatic representation of the electron self-energy $\Sigma=g^2GD_0$, where $g$ are the screened EPC matrix elements (grey circles), $G$ is the electron Green's function (straight line) and $D_0$ is the bare phonon propagator (single wavy line). (b) The electron self-energy $\Sigma=g^2GD$, where the bare phonon propagator $D_0$ is replaced with the dynamical (NA) phonon propagator $D$ (double wavy line). (c) Dressing of the phonons via the dynamical EPC, i.e., the Dyson equation for the phonon propagator $D=D_0+D_0 \widetilde{\pi} D$, where $\widetilde{\pi}$ is the NA phonon self-energy due to EPC.
} 
\label{fig:fig1}
\end{figure}

\section{Results}

A central quantity in many-body perturbation theory for describing electron dynamics due to EPC is the electron self-energy $\Sigma$, which is instrumental for understanding quasiparticle spectral features\,\cite{damascelli03,duvel22}, superconductivity properties, such as $T_c$\,\cite{allen83,marsiglio08}, electron scattering rates, and electron resistivity\,\cite{allen74}. Schematically, the electron self-energy can be written as $\Sigma=g^2GD$, where $g$ are the screened EPC matrix elements, $G$ is the electron Green's function and $D$ is the phonon propagator [see diagram in Fig.\,\ref{fig:fig1}(a)]. In the standard calculations of $\Sigma$ based, e.g., on DFPT, the phonon propagator $D$ is non-interacting with sharp (i.e., having infinitive lifetime) adiabatic phonons\,\cite{baroni01,giustino17,epw}. In the following we inspect how general electron-phonon properties are modified when dynamical properties are accounted for in $D$ via NA EPC, as diagrammatically depicted in Figs.\,\ref{fig:fig1}(b) and \ref{fig:fig1}(c), where additional interaction between phonons and electron-hole pairs brings about finite phonon lifetime and frequency renormalizations.

\begin{figure}[!b]
\includegraphics[width=0.45\textwidth]{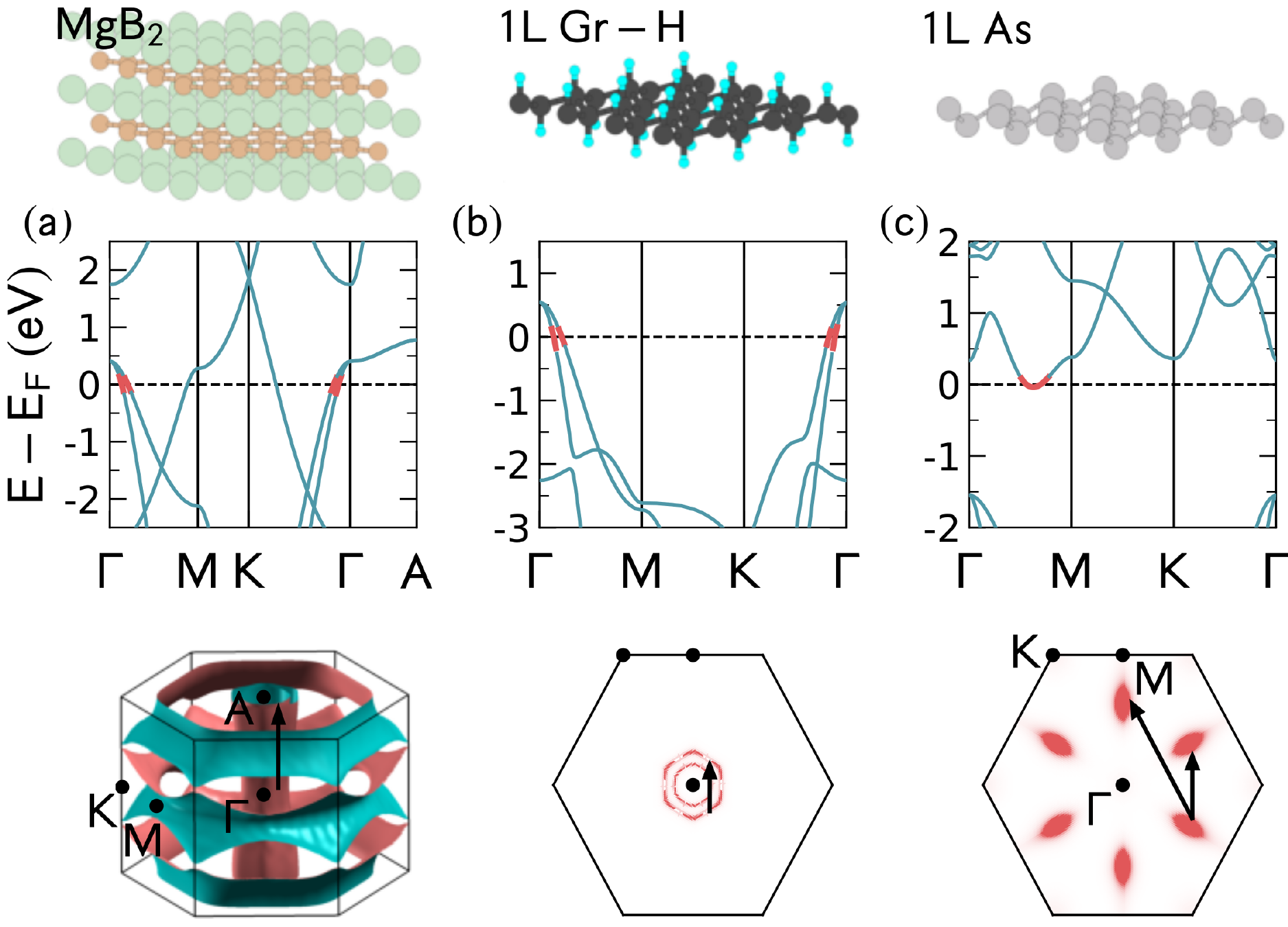}
\caption{Structure, electronic bands along high-symmetry points of the Brillouin zone, and Fermi surface in the first Brillouin zone for (a) MgB$_2$, (b) graphane (1L Gr-H), and (c) arsenene (1L As). Red/green color highlights the relevant electronic states at the Fermi level that are involved in electron-phonon scatterings and formation of the (adiabatic and dynamical) Kohn anomalies. Black arrows depict the corresponding electronic transitions.
} 
\label{fig:fig2}
\end{figure}
\begin{figure*}[!t]
\includegraphics[width=0.95\textwidth]{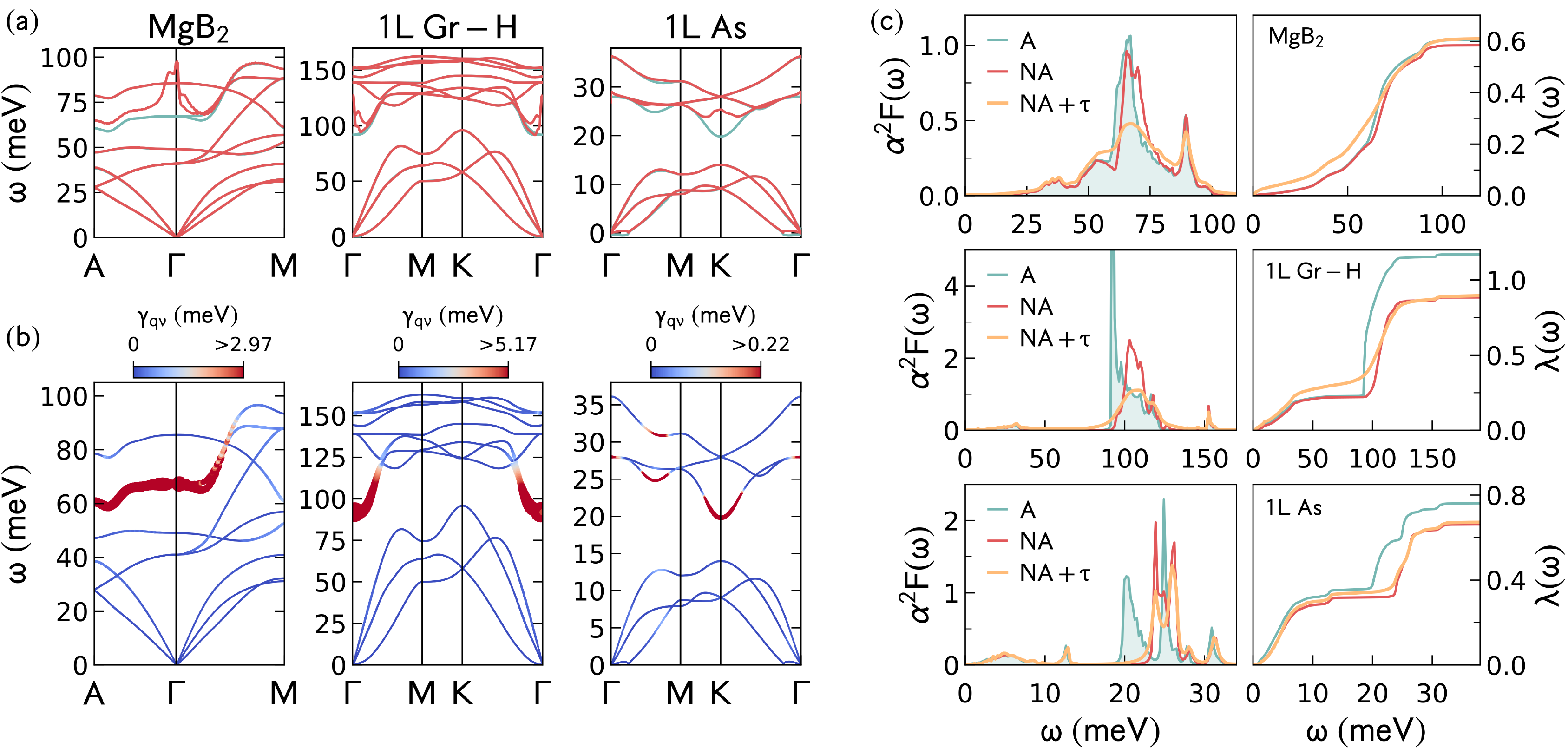}
\caption{(a) Adiabatic (A; blue) and nonadiabatic (NA; red) phonon dispersions of MgB$_2$, monolayer graphane (1L Gr-H), and monolayer arsenene (1L As). Notice how strong dynamical corrections to the phonon bands can occur away from the center of the Brillouin zone, as it is the case in MgB$_2$ and 1L As. (b) The corresponding phonon linewidths $\gamma_{\mathbf{q}\nu}$ coming from the EPC. The size of the dots and colors (defined within the colorbars) represents the intensity of $\gamma_{\mathbf{q}\nu}$. (c) Eliashberg functions $\alpha^2F(\omega)$ and cumulative EPC constant $\lambda(\omega)$ obtained for MgB$_2$, 1L Gr-H, and 1L As. Three different results are shown: adiabatic (A), where frequencies are obtained within the adiabatic DFPT and without momentum and branch-resolved phonon broadenings $\gamma_{\mathbf{q}\nu}$ (blue), nonadiabatic (NA), where frequencies are corrected with NA effects, while the corresponding NA phonon broadening due to EPC is not included (red), and full nonadiabatic (NA+$\tau$), where both NA frequency renormalization and NA phonon linewidth effects are taken into account (yellow).
} 
\label{fig:fig3}
\end{figure*}

When averaged over the Fermi surface, the electron self-energy $\Sigma$ is commonly expressed via the electron-phonon spectral function or the Eliashberg function\,\cite{allen74,allen83}
\begin{equation}
\alpha^2F(\omega)=\frac{1}{\pi N(E_F)}\sum_{\mathbf{q}\nu}\frac{\gamma_{\mathbf{q}\nu}}{\Omega_{\mathbf{q}\nu}}B_{\nu}(\mathbf{q},\omega),
\label{eq:eq1}
\end{equation}
where $\mathbf{q}$ and $\nu$ are phonon momentum and branch index, $N(E_F)$ is the density of states at the Fermi level, $\gamma_{\mathbf{q}\nu}$ are the phonon linewidths in the double-delta approximation\,\cite{allen72}, $\Omega_{\mathbf{q}\nu}$ are the phonon frequencies, while $B_{\nu}(\mathbf{q},\omega)$ is the phonon spectral function. If one accounts for the dynamical EPC effects, the latter quantities can be expressed via the NA phonon self-energy $\widetilde{\pi}_{\nu}(\mathbf{q},\omega)$, i.e., $\gamma_{\mathbf{q}\nu}=-{\rm Im}\,\widetilde{\pi}_{\nu}(\mathbf{q},\Omega_{\mathbf{q}\nu})$, $\Omega^2_{\mathbf{q}\nu}=\omega^2_{\mathbf{q}\nu}+2\omega_{\mathbf{q}\nu}{\rm Re}\,\widetilde{\pi}_{\nu}(\mathbf{q},\Omega_{\mathbf{q}\nu})$, and
\begin{equation}
B_{\nu}(\mathbf{q},\omega)=-\frac{1}{\pi}\mathrm{Im}\left[ \frac{2\omega_{\mathbf{q}\nu}}{\omega^2-\omega_{\mathbf{q}\nu}^2-2\omega_{\mathbf{q}\nu}\widetilde{\pi}_{\nu}(\mathbf{q},\omega)} \right].
\label{eq:eq2}
\end{equation}
Note that in the standard adiabatic simulations of the EPC properties, the dynamical effects are absent, and one calculates Eqs.\,\eqref{eq:eq1} and \eqref{eq:eq2} with $\widetilde{\pi}_{\nu}(\mathbf{q},\omega)\rightarrow i0$\,\cite{giustino17,epw}. Here we calculate NA phonon self-energy as $\widetilde{\pi}_{\nu}(\mathbf{q},\omega)=\pi_{\nu}(\mathbf{q},\omega)-\pi_{\nu}(\mathbf{q},0)$ [since the adiabatic frequency $\omega_{\mathbf{q}\nu}$ already contains $\pi_{\nu}(\mathbf{q},0)$], where
\begin{equation}
\pi_{\nu}(\mathbf{q},\omega)=\sum_{\mathbf{k}nm}\left| g_{\nu}^{nm}(\mathbf{k},\mathbf{q}) \right|^2\frac{f_{n\mathbf{k}}-f_{m\mathbf{k+q}}}{\omega+\varepsilon_{n\mathbf{k}}-\varepsilon_{n\mathbf{k+q}}+i\eta}.
\label{eq:eq3}
\end{equation}
The electron-phonon matrix elements are denoted with $g_{\nu}^{nm}(\mathbf{k},\mathbf{q})$, Fermi-Dirac distribution functions are $f_{n\mathbf{k}}$, $\varepsilon_{n\mathbf{k}}$ are electron energies, and $\eta$ is an infinitesimal parameter. Equations \eqref{eq:eq1}-\eqref{eq:eq3} and the corresponding input parameters are calculated in this work by means of DFPT\,\cite{baroni01,qe} and Wannier interpolation\,\cite{wan90} of EPC matrix elements $g_{\nu}^{nm}$\,\cite{epw}. All the necessary computational details can be found in Ref.\,\cite{SM}.

In Fig.\,\ref{fig:fig2} we show the electronic structures of MgB$_2$, hole-doped 1L Gr-H, and electron-doped 1L As. The electronic band structure of MgB$_2$ around the Fermi level consists of the hole-like $\sigma$ states around the center, and $\pi$ states at the edges of the Brillouin zone. The former (red thicker lines) are the most relevant for the formation of the Kohn anomalies in the phonon spectra and provide a dominating contribution to the superconducting state. In 1L Gr-H the Fermi surface consists only of the hole $\sigma$ states around the $\Gamma$ point, while in 1L As there are 6 electron pockets appearing between the $\Gamma$ and M points. All these states are involved in the strong electron-phonon scatterings (see arrows in Fig.\,\ref{fig:fig1}), in these cases mostly with optical phonons, and are therefore an important contribution to the total electron-phonon coupling strength $\lambda$ and superconducting properties. Band structures of other systems considered in this work (C, 1L Gr, 1L MoS$_2$, 1L InSe, and 1L WTe$_2$) are provided in Fig.\,S1\,\cite{SM}. 

Calculations of the phonon dispersions and EPC strengths are outlined in Fig.\,\ref{fig:fig3}, where standard adiabatic and dynamical results are compared. Interestingly, along with the expected NA frequency renormalizations at the $\Gamma$ point for the systems with the strong EPC (1L Gr-H), there are considerable NA corrections of phonons away from the center of the Brillouin zone for MgB$_2$ and 1L As. As shown in Fig.\,S2\,\cite{SM}, 1L MoS$_2$, 1L InSe, and 1L WTe$_2$ are also characterized with strong dynamical Kohn anomalies away from the $\Gamma$ point.
Therefore, the NA condition $\mathbf{q\cdot v_F}<\omega$\,\cite{engelsberg63,maksimov96} only holds for simple metals with parabolic band structure, while for the multiband systems, strong electron-phonon scatterings between same or different valleys could lead to NA Kohn anomalies at finite $\mathbf{q}=\mathbf{q_c}$, provided that the $\varepsilon_{n\mathbf{k}} -\varepsilon_{m\mathbf{k+q_c}} \lesssim \omega$ condition is met. These $\mathbf{q}=\mathbf{q_c}$ dynamical transitions are shown with arrows in Fig.\,\ref{fig:fig2} for MgB$_2$, 1L Gr-H, and 1L As, and represent the largest contributions to the electron-phonon scatterings and total EPC strength $\lambda$. For instance, in the case of MgB$_2$ strong NA corrections are present for the optical $E_{2g}$ mode around the $\Gamma$ point, but also extending along the $\mathrm{\Gamma}-\mathrm{M}$ line, exactly where the phonon linewidths $\gamma_{\mathbf{q}\nu}$ are largest. It is similar in other cases, where intra- (C, 1L Gr-H, 1L Gr, 1L InSe) and inter-valley (1L As, 1L MoS$_2$, 1L WTe$_2$) electron-phonon scatterings bring about, at the same wavevector $\mathbf{q_c}$ and phonon branch, strong NA corrections to phonon frequencies and large NA phonon linewidths $\gamma_{\mathbf{q}\nu}$. Notice that the conditions that lead to the finite-$\mathbf{q}$ NA Kohn anomalies are similar to the criteria for the phonon-induced CDW formation\,\cite{johannes06}, which include considerable nesting features of the Fermi surface or strong EPC matrix elements for $\mathbf{q}=\mathbf{q_c}$.
In accordance to these results, the NA hardening of the soft phonon modes at the edges of the Brillouin zone were obtained recently also for TaS$_2$\,\cite{berges22}.

\begin{figure}[!b]
\includegraphics[width=0.45\textwidth]{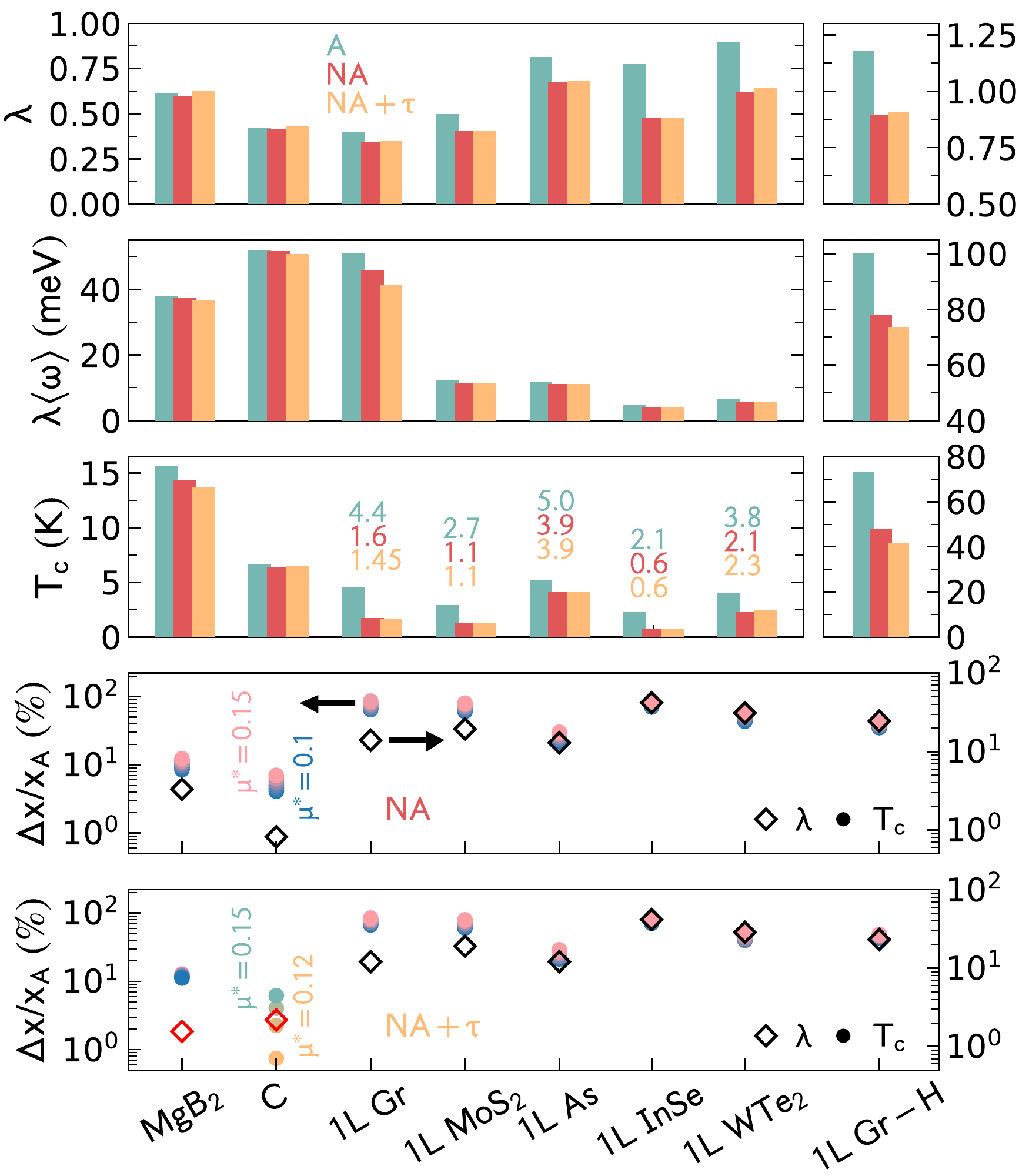}
\caption{The total EPC strengths $\lambda$, the first moments of the phonon spectrum $\lambda\langle\omega\rangle$, and superconducting transition temperatures $T_c$ obtained with three different approaches: adiabatic (A; within standard adiabatic DFPT methodology), nonadiabatic (NA; by including NA phonon frequency renormalizations), and full nonadiabatic (NA+$\tau$; by including both NA frequency renormalizations and NA phonon broadenings $\gamma_{\mathbf{q}\nu}$). The results are presented for various bulk an 2D superconducting materials.
Bottom panels show the corresponding relative changes $\mathrm{(x_{NA/NA+\tau}-x_A)/x_A}$ for $T_c$ (left axis) and $\lambda$ (right axis). All relative changes are positive, except the NA+$\tau$ results of $\mathrm{x=\lambda}$ for MgB$_2$ and C (red diamond signs), and of $\mathrm{x}=T_c$ for C (green-yellow circles).
} 
\label{fig:fig4}
\end{figure}

The aforementioned two NA effects are often inseparable and should be included together when calculating the Eliashberg function $\alpha^2 F(\omega)$ and the total EPC strength $\lambda$. In Fig.\,\ref{fig:fig3}(c) we show the results of $\alpha^2 F(\omega)$ and cumulative EPC constant $\lambda(\omega)=2\int_0^{\omega}d\Omega \alpha^2F(\Omega)/\Omega$ for three different approaches. Namely, we use the standard adiabatic approximation, where no dynamical effects are considered in Eq.\,\ref{eq:eq1}, i.e., $\widetilde{\pi}_{\nu}(\mathbf{q},\omega)\rightarrow i0$, (blue), the NA approach with only dynamical frequency renormalizations included, i.e., $\widetilde{\pi}_{\nu}(\mathbf{q},\omega)\rightarrow {\rm Re}\widetilde{\pi}_{\nu}(\mathbf{q},\omega)+i0$, (red), and the full NA method with both frequency corrections and phonon dampings (yellow). The NA frequency renormalizations are always positive\,\cite{giustino17}, and since these shifts are accompanied by strong phonon damping rates, the main peaks in $\alpha^2 F(\omega)$ are blueshifted and $\lambda$ is reduced. For example, the NA blueshifts of the Kohn anomalies of about 33\,meV and 6\,meV in 1L Gr-H and 1L As, respectively, is reflected in the corresponding modifications in $\alpha^2 F(\omega)$ and in a considerable reduction of $\lambda$ of about 0.29 and 0.14. By including additionally the momentum- and mode-resolved broadening $\gamma_{\mathbf{q}\nu}$, the total EPC constant $\lambda$ is not affected seriously, however, the spectral weight of $\alpha^2 F(\omega)$ is redistributed and smoothed, so that the high-frequency main peaks contribute less, while lower frequencies contribute more to $\lambda(\omega)$. This frequency redistribution of the EPC due to NA effects has pertinent consequences on the low-temperature and low-energy behavior of electron-hole scattering rate\,\cite{allen74}, as we will show below. For more data on $\alpha^2F(\omega)$ see Fig.\,S3\,\cite{SM}.

\begin{figure*}[!t]
\includegraphics[width=0.95\textwidth]{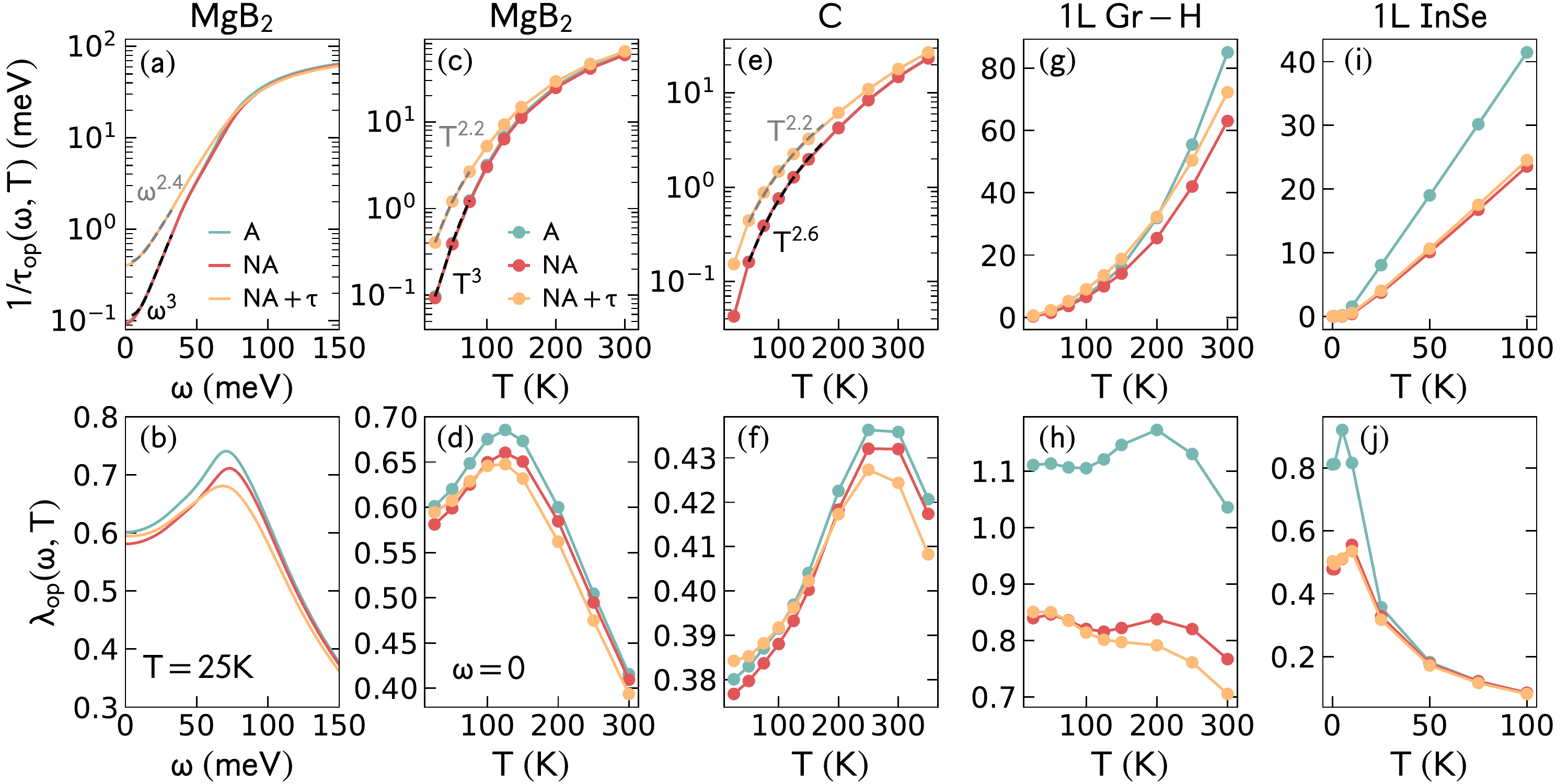}
\caption{Frequency and temperature dependence of the optical (or electron-hole pair) scattering rate $1/\tau_{\rm op}(\omega,T)$ and the mass enhancement (or energy renormalization) parameter $\lambda_{\rm op}(\omega,T)$ for various bulk and 2D materials as obtained with adiabatic (A), nonadiabatic (NA), and full nonadiabatic (NA+$\tau$) methods. Note how the scaling of the low-frequency and low-temperature regime of $1/\tau_{\rm op}$ is modified under the effect of NA phonon broadening.
} 
\label{fig:fig5}
\end{figure*}

Collected results for all of the studied systems is presented in Fig.\,\ref{fig:fig4}, including the dynamical corrections of the total EPC strength $\lambda$, the first moment of the phonon spectrum $\lambda\langle\omega\rangle$, and the superconducting transition temperature $T_c$. The first moment is calculated as $\lambda\langle\omega\rangle=2\int_0^{\infty}d\Omega \alpha^2F(\Omega)$, and it is considered important for the McMillan's expression for $T_c$\,\cite{mcmillan89}, as well as high-energy and zero-temperature estimation of the electron-hole relaxation rate $1/\tau_{\rm op}$ appearing in the optical conductivity formula\,\cite{allen71,shulga91,marsiglio08,novko17}. For calculations of $T_c$ we use Allen-Dynes version of McMillan's formula\,\cite{allen75}. While in some cases the dynamical renormalizations of frequencies induce small relative modifications (compared to adiabatic result) of $\lambda$ (MgB$_2$, C), there are cases where it is significantly decreased, i.e., by $19\%$ to $42\%$ (MoS$_2$, 1L InSe, 1L WTe$_2$, and 1L Gr-H). This is reflected in mild and considerable reduction in $T_c$, respectively. For example, 8-12$\%$ for MgB$_2$ and  70-85$\%$ for InSe (as obtained with reasonable choices of effective Coulomb repulsion $\mu^{\ast}$). In most cases the dynamical effects on $\lambda\langle\omega\rangle$ are small (largest being for 1L Gr-H), suggesting that high-energy limits of $1/\tau_{\rm op}$ and, consequently, phonon-assisted optical absorption formula are not strongly affected by the NA renormalizations. Interestingly, the value of $\lambda$ can be additionally increased by the NA phonon broadening effects, while $T_c$ can be either increased or decreased, depending on an interplay between modifications in the position of strong peaks in $\alpha^2F(\omega)$ (via $\langle\omega\rangle$ or $\omega_{\rm log}$) and changes in $\lambda$. For instance, the phonon-broadening-induced reduction in $\omega_{\rm log}$ (see Fig.\,S4\,\cite{SM}) and increase in $\lambda$ results in further decrease of $T_c$ for MgB$_2$, 1L Gr, and 1L Gr-H. On the other hand, the NA phonon broadening and the accompanying redistribution of EPC in $\alpha^2F(\omega)$ induce small increase of $T_c$ in C (e.g., around 6$\%$ for $\mu^{\ast}=0.15$). Note also a small broadening-induced increase of $T_c$ in the case of 1L WTe$_2$. The latter two examples serve as an illustration on how the NA broadening effects can be quite beneficial for obtaining higher values of $T_c$, and it is quite possible that these dynamical corrections could be higher and more emphasized for systems with soft phonon modes\,\cite{kaga22,jiang22} that are anharmonic and strongly coupled to electrons\,\cite{setty20,setty22} (e.g., charge-density-waves materials and systems at the verge of phase transition).

Having in mind the available experimental and theoretical adiabatic results, we emphasize that the NA renormalizations presented here generally improve the agreement. Namely, the theoretical first-principles estimations of $T_c$ for MgB$_2$ are 50-55\,K when anisotropic electron-phonon interaction and averaged Coulomb repulsion term are employed\,\cite{choi02b,floris05,margine13}, while experimental value is $T_c=39$\,K\,\cite{choi02} (difference of about 22-29$\%$). As we show here by using the isotropic Allen-Dynes solution to Eliashberg equations, the substantial part of this discrepancy could be resolved by the NA effects (inducing decrease of $T_c$ by 8-12$\%$). Additional improvements might be achieved by accounting for vertex corrections in the Eliashberg equations\,\cite{cappelluti02} and anharmonic effects\,\cite{choi02b}.
Further, the \emph{ab-initio} results of $T_c$ for the hole-doped diamond are underestimating the experimental values\,\cite{boeri04,giustino07}, while the present results demonstrate how the NA phonon broadening in C can enhance $T_c$. However, we note that the obtained increase is not enough to reproduce the experiments and that the boron-related vibrational modes are instrumental for understanding the superconductivity in hole-doped diamond\,\cite{xiang04,giustino07b}. Also, regardless of the doping concentration and the type of calculation, theoretical predictions of transition temperature in 1L MoS$_2$ ($T_c\gtrsim 4$\,K)\,\cite{ge13,rosner14,goiricelaya19} are always much higher than experimental values, which are $\sim $1-2\,K\,\cite{costanzo16,fu17}. Our results suggest that the dynamical renormalizations could improve this disagreement. Experimental results for WTe$_2$ are available as well, however, only for small electron concentrations\,\cite{ebrahim18,yang20}, where we do not expect significant NA modifications.

We would also like to point out that the previous adiabatic results of $\lambda$ and $T_c$ for all of the presented systems are in a good agreement with our adiabatic results (see Fig.\,S4\,\cite{SM}).

Recently, dynamical effects on superconducting properties via dynamical screening of EPC matrix elements were studied by means of real-time time-dependent-density-functional theory combined with molecular dynamics\,\cite{meng22a}.
The corresponding conclusion, based on the 1L Gr and 1L Gr-H cases, is that the dynamical screening that enters $g^2$, e.g., via the dielectric function $\epsilon(\mathbf{q},\omega)$, can overcome the effects of the NA frequency renormalization (hardening), and lead to the enhancement of $\lambda$ and $T_c$ compared to the static-screening calculations. These results are, however, hard to compare with the present work. On the one hand, the dynamical screening of the EPC matrix elements $g$ is unfortunately out of reach for the present many-body perturbation approach that uses phonon self-energy within DFPT\,\cite{calandra10,giustino17,harkonen20,marini22}. On the other hand, the approach from Ref.\,\cite{meng22a} includes the dynamically-screened $g$ and NA frequency renromalizations, but, does not account for the NA phonon broadenings. Also, it uses a frozen-phonon scheme to account for phonon dynamics and is therefore able to include only a few modes in calculations of the total $\lambda$ (i.e., the $\mathbf{q}=0$ and $\mathbf{q}=K$ optical modes for 1L Gr, and only the $\mathbf{q}=0$ optical mode for 1L Gr-H). We stress out that the fine $\mathbf{q}$ grids are necessary to reach the numerical convergance and account for the full NA effects of the Kohn anomalies. Also, we do not expect that the dynamical screening of EPC will play an important role in modifying the EPC properties for the materials with NA Kohn anomalies away from the $\Gamma$ point, e.g., in multivalley materials such as 1L MoS$_2$, 1L As, and 1L WTe$_2$, since $\epsilon(\mathbf{q},\omega)\approx \epsilon(\mathbf{q},0)$ for large $\mathbf{q}$\,\cite{verdi17,duvel22}. Nonetheless, the dynamical screening of $g$ seems to be quite important feature of the NA theory, and it might be a crucial (but still very demanding) extension of the present DFT-based many-body approach, e.g., to explore the impact of the strong plasmon-phonon coupling on the superconductivity and transport\,\cite{bauer09,krsnik22}.

Figure \ref{fig:fig5} depicts the results for optical (i.e., electron-hole pair) scattering rate $1/\tau_{\rm op}$ that enters optical conductivity formula (or current-current response tensor) and can be expressed in the following form\,\cite{allen74,shulga91,marsiglio08,novko2018a,novko20b,novko2020a}
\begin{align}
1/\tau_{\rm op}(\omega,T)=\frac{\pi}{\omega} \int d\Omega \alpha^2F(\Omega)\Big[2\omega \coth\frac{\Omega}{2k_BT}\nonumber\\-(\omega+\Omega)\coth\frac{\omega+\Omega}{2k_BT}+(\omega-\Omega)\coth\frac{\omega-\Omega}{2k_BT}\Big],
\label{eq:eq4}
\end{align}
having frequency $\omega$ and temperature $T$ dependence. In addition, the results of the concomitant mass enhancement parameter (i.e., energy renormalization of electron-hole pairs) $\lambda_{\rm op}(\omega,T)$ are shown. The latter quantity is obtained by performing Kramers-Kronig transormation of Eq.\,\eqref{eq:eq4}. Both $1/\tau_{\rm op}$ and $\lambda_{\rm op}$ are essential in understanding the temperature dependence of optical conductivity\,\cite{allen74}, plasmon\,\cite{kupcic15,novko17} and phonon\,\cite{novko17} damping (when $\omega\neq 0$), as well as for resistivity\,\cite{allen71,chakraborty76}, band renormalization\,\cite{grimvall69,allen70}, and specific heat\,\cite{grimvall69,allen74} (when $\omega=0$). Note also that the EPC constant is $\lambda=\lambda_{\rm op}(0,0)$.

The results show intriguing modifications of the scaling law for small-$\omega$ and small-$T$ regime of scattering rate $1/\tau_{\rm op}$. Namely, the NA phonon broadening effects that enter the phonon spectral function can significantly decrease the exponents $x$ and $y$ in $1/\tau_{\rm op}\propto a\omega^x+bT^y$, generally, from about 3 to 2 (see Fig.\,S4 in Ref.\,\cite{SM}). This NA effect is particularly emphasized for the cases where the optical phonons with large phonon linewidths are present, e.g., for MgB$_2$ and C. Then the large phonon linewidths extend to the low-$\omega$ part and modifies the corresponding region of $\alpha^2F(\omega)$. 
It is interesting to note that even though the value of the measured resistivity for MgB$_2$ varies between different samples, it usually shows roughly $T^2$ dependence at lower temperatures\,\cite{rowell03}, potentially confirming our predictions.
The scaling change is even more pronounced for systems with low-$\omega$ acoustic modes at finite $\mathbf{q}$ having strong NA broadening, e.g., as it is the case for 1L MoS$_2$ and 1L InSe. The modification of the scaling law for $1/\tau_{\rm op}$ from the usual $T^3$ to $T^2$ that comes from the self-consistent treatment of the NA EPC was already discussed $\sim50$ years ago\,\cite{fenton69,allen74}, but it was never thoroughly appreciated, examined and applied to the real materials. Here we confirm that the $T^2$ and $\omega^2$ dependence of $1/\tau_{\rm op}$ does not necessarily refer to the Fermi liquid result for the electron-electron scattering, but it can equally be a fingerprint of the strong dynamical electron-phonon scattering. By combining the diagrams presented in Figs.\,\ref{fig:fig1}(b) and \ref{fig:fig1}(c), it is obvious that one can look at the present NA effects in $1/\tau_{\rm op}$ as the effective electron-electron scatterings mediated by phonons, and thus the same phase-space arguments for the Fermi liquid result apply here.
Note also the substantial modifications of $1/\tau_{\rm op}$ at higher (i.e., linear) $T$ that come from the large modifications of EPC $\lambda$ due to NA frequency renomralizations, like in the case of 1L Gr-H and 1L InSe.

The NA effects are as well reflected in functional and intensity modifications of mass enhancement parameter $\lambda_{\rm op}(\omega,T)$. Note for instance, that the $\omega$ and $T$ scaling laws are also modified in $\lambda_{\rm op}$ for small $\omega$ and $T$ when NA phonon broadening effects are accounted for. Large NA-induced modifications of $\lambda$, i.e., $\lambda_{\rm op}(0,0)$, are extended to finite-$T$ region, as it is evident for 1L Gr-H and 1L InSe.
This should, for example, have direct implications on the electron effective mass and its temperature dependence $m^{\ast}=m_e [1+\lambda_{\rm op}(0,T)]$, i.e., the band dispersion renormalizations around the Fermi level, and its microscopic \emph{ab-initio} explanations, which are usually rooted in the adiabatic theory\,\cite{allen13,mechelen08,marel11,zacharias20}. Further, finite temperature $\lambda_{\rm op}(0,T)$ [as well as $1/\tau_{\rm op}(0,T)$], renormalized by the NA effects, which enters electron self-energy in energy-gap Eliashberg equations, can have some interesting impact on estimations of high-temperature superconductivity\,\cite{appel68,pickett80} (notice that such thermal-phonon effects are absent in McMillan's formula for $T_c$).

\section{Discussion}

In summary, we have thoroughly examined the influence of the NA phonon dynamics on electron-phonon properties in several theoretically and experimentally established superconductors. Contrary to the common belief, we demonstrate that the effect of ``slow'' electrons that lag behind the ``fast'' moving atoms is actually quite common and it should be considered not only for understanding the dynamics of electrons\,\cite{ponce15,miglio20}, but also for phonon dynamics and consequently for many physical phenomena that are founded in the electron-phonon interaction. In fact, it was recently presented in a compelling way, that nonadiabaticity is linked with the Drude weight of conduction electrons\,\cite{dreyer22}, which is always finite for metals and doped semiconductors.
The results in Refs.\,\cite{novko20b,sohier19} show that the most dramatic dynamical modifications will occur when phonon-induced perturbations of conduction electron density in adiabatic and nonadiabatic regimes are disparate as well as associated with the strong EPC. See for instance Fig.\,3 in Ref.\,\cite{novko20b} for the nonadiabatic perturbation of charge density in the case of MoS$_2$ optical out-of-plane phonon at $\mathbf{q}=0$. Here we show that the same applies for Kohn anomalies away from the Brillouin zone center, with serious repercussions on the total EPC constant $\lambda$ and superconducting properties.

We thus believe that the NA-induced modifications of EPC are quite universal and that
an extension of the present NA treatment might be applicable and useful for explaining electron-phonon properties in a number of unusual and strongly-correlated materials hosting CDW, soft phonon modes or Kohn anomalies, phase transitions, superconducting phase, and enigmatic $T^2$ resistivity. For instance, the latter scaling law of resistivity was observed concomitantly in materials characterized with the high-$T_c$ superconducting phase (e.g., cuprates\,\cite{barisic13}), Mott transition (e.g., SrVO$_3$\,\cite{mirjolet21}), structural phase transition and superconductivity (e.g., SrTiO$_3$\,\cite{marel11,lin15}), as well as in more conventional systems like Al\,\cite{macdonald80} and TiS$_2$\,\cite{thompson75}. Recently, the $T^2$ law was observed and discussed for novel graphene-based heterostructures, like graphene superlattice on hBN\,\cite{wallbank18} and twisted bilayer graphene\,\cite{jaoui22}. The superconductivity phase was discovered in the latter case, where the strongly coupled and damped optical phonons might have an important contribution to the total EPC\,\cite{choi18,choi21}.
In all of these cases the corresponding interpretations were typically given in terms of the Fermi liquid theory ($1/\tau_{\rm FL}\propto T^2$), i.e., electron-electron scattering, or some corresponding deviations. On the other hand, here we confirm Allen's result\,\cite{allen74} that $T^2$ scaling is not unique to the Fermi liquid result for the scattering rate, and it could also be a sign for the presence of the strongly damped phonons and enhanced EPC. Similar speculations were provided by Feton \emph{et al.}\,\cite{fenton69}, as well as by MacDonald\,\cite{macdonald80}.

It would be interesting to examine if the NA renormalizations affect the superconducting properties in hydride systems at high pressure\,\cite{floreslivas20}, where first-principles prediction of $T_c$ are sometimes overestimating the experimental values\,\cite{errea16} and some signatures of Kohn anomalies in calculated phonon spectra are present\,\cite{errea15}. As well as for proposed few-layer and bulk derivatives of MgB$_2$, where superconductivity was theoretically studied\,\cite{bekaert19,singh22,yu22}.

Furthermore, notice that the NA corrections might be important for precise determination of transition temperature of CDW and structural phase transition\,\cite{alidoosti21,alidoosti22,novko22,berges22}, since we show that both $\mathbf{q}\approx0$ and $\mathbf{q}>0$ Kohn anomalies are often strongly affected by the dynamical renormalizations, as well as for the electron-phonon part of the thermal conductivity\,\cite{kim16}.

\begin{acknowledgments}
Useful discussions with Jan Berges, Samuel Ponc\'e, Tim Wehling, Osor S. Bari\v{s}i\'c, and Juraj Krsnik are gratefully acknowledged. We acknowledge financial support from the Croatian Science Foundation (Grant no. UIP-2019-04-6869) and from the European Regional Development Fund for the ``Center of Excellence for Advanced Materials and Sensing Devices'' (Grant No. KK.01.1.1.01.0001).
\end{acknowledgments}

\bibliography{ref}

\end{document}